\documentclass[pra,twocolumn,showpacs,floatfix]{revtex4}
\usepackage{graphicx}

\newcommand{\beq}{\begin{equation}}
\newcommand{\eeq}{\end{equation}}
\newcommand{\beqa}{\begin{eqnarray}}
\newcommand{\eeqa}{\end{eqnarray}}
\newcommand{\ba}{\begin{array}}
\newcommand{\ea}{\end{array}}

\begin{document}

\title{Effects of Axial Vorticity in Elongated Mixtures of Bose-Einstein
Condensates}
\author{L. Salasnich$^{1}$, B. A. Malomed$^{2}$ and F. Toigo$^{1}$}
\affiliation{$^1$CNISM and CNR-INFM, Unit\`a di Padova, \\
Dipartimento di Fisica ``Galileo Galilei'', Universit\`a di Padova, Via
Marzolo 8, 35131 Padova, Italy \\
$^{2}$Department of Physical Electronics, School of Electrical Engineering,
Faculty of Engineering, Tel Aviv University, Tel Aviv 69978, Israel }

\begin{abstract}
We consider a meniscus between rotating and nonrotating species in the
Bose-Einstein condensate (BEC) with repulsive inter-atomic interactions,
confined to a pipe-shaped trap. In this setting, we derive a system of
coupled one-dimensional (1D) nonpolynomial Schr\"{o}dinger equations (NPSEs)
for two mean-field wave functions. Using these equations, we analyze the
phase separation/mixing in the pipe with periodic axial boundary conditions,
i.e. in a toroidal configuration. We find that the onset of the mixing, in
the form of \textit{suction}, i.e., filling the empty core in the vortical
component by its nonrotating counterpart, crucially depends on the vorticity
of the first component, and on the strengths of the inter-atomic
interactions.
\end{abstract}

\maketitle

Since the creation of vortices in Bose-Einstein condensates (BECs)\ \cite%
{Cornell1,Cornell2(filled-empty)}, this topic has been a subject of many
experimental and theoretical works, as reviewed in Refs. \cite{reviews}. In
particular, much attention has been drawn to vortices in mixtures of two BEC
species; in fact, the first vortices were created in a two-component setting
\cite{Cornell1}, and a theoretical analysis of that setting was developed
too \cite{multicomp}. In this connection, a situation of straightforward
interest is the interaction of rotating and nonrotating immiscible BEC
species. It is natural to expect that the nonrotating component may fill the
hollow vortical core(s) in the rotating one. As shown experimentally,
vortices and vortex lattices with empty and filled cores feature a great
difference in their structure \cite{Cornell2(filled-empty),Cornell3}. Very
recently it has been also predicted the possibility to create vortex with
arbitrary topological charge by phase engineering \cite{finland}.

Matter-wave vortices can be created not only in large-aspect ratio settings,
but also in narrow cigar-shaped traps (\textquotedblleft pipes"), which help
to stabilize them \cite{cigar-vortex}. In the pipe geometry, it is
interesting too to consider the interaction of rotating and nonrotating BEC
species, which is the subject of the present work. In particular, a natural
issue in this case is the effect of \textquotedblleft suction", i.e., onset
of effective mixing between two nominally immiscible species by pushing the
nonrotating component into the empty core in the vorticity-carrying one. In
other words, the suction implies indefinite stretching of the meniscus
separating the immiscible species which originally fill two halves of the
tube.

In the mean-field approximation, the starting point of the analysis is a
system of coupled Gross-Pitaevskii equations (GPEs) for macroscopic wave
functions, $\psi _{1}$ and $\psi _{2}$, of the two BEC species confined in
the tight cylindrical trap. In the scaled form, the equations are%
\begin{eqnarray}
i\frac{\partial \psi _{1}}{\partial t} &=&\left[ -\frac{1}{2}\nabla ^{2}+%
\frac{1}{2}\left( x^{2}+y^{2}\right) \right.  \nonumber \\
&&\left. +\left( g_{1}|\psi _{1}|^{2}+g_{12}\left\vert \psi _{2}\right\vert
^{2}\right) \right] \psi _{1},  \label{psi}
\end{eqnarray}%
\begin{eqnarray}
i\frac{\partial \psi _{2}}{\partial t} &=&\left[ -\frac{1}{2m}\nabla ^{2}+%
\frac{1}{2}\Omega ^{2}\left( x^{2}+y^{2}\right) \right.  \nonumber \\
&&\left. +\left( g_{2}\left\vert \psi _{2}\right\vert ^{2}+g_{12}|\psi
_{1}|^{2}\right) \right] \psi _{2},  \label{phi}
\end{eqnarray}%
where $z$ is the axial coordinate, while $x$ and $y$ are the transverse
ones. In these equations, lengths are measured in units of $a_{\bot }^{(1)}=%
\sqrt{\hbar /(m_{1}\Omega _{1})}$, the energy in units of $\hbar \Omega _{1}$%
, and time in units of $1/\Omega _{1}$, where $m_{1,2}$ and $\Omega _{1,2}$
are masses and transverse trapping frequencies of the two species, while the
relative parameters in Eqs. (\ref{psi}) and (\ref{phi}) are $m\equiv
m_{2}/m_{1}$ and $\Omega \equiv \Omega _{2}/\Omega _{1}$. The interaction
strengths in the equations are expressed in terms of the $s$-wave
inter-atomic scattering lengths, $g_{1,2}\equiv 2a_{1,2}/a_{\bot }^{(1)}$, $%
g_{12}\equiv 2a_{12}/a_{\bot }^{(1)}$, where $a_{j}$ is the scattering
length in the $j$-th species, while $a_{12}$ is the scattering length for
atoms belonging to the different species. In this work, we consider the most
relevant situation with the repulsion between atoms belonging to the same
and different species, i.e., we take $g_{1},g_{2},g_{12}>0$.

We will solve equations (\ref{psi}) and (\ref{phi}), as well as effective 1D
equations to be derived from them, with periodic boundary conditions in the
axial direction ($z$), which assumed that the pipe is closed into a torus of
a large radius. Toroidal traps of the magnetic type are currently available
to the experiment \cite{torus-experiment}, and the BEC\ dynamics in the
toroidal geometry was studied theoretically in several works \cite%
{torus-theory}. Actually, a variety of the toroidal configuration is a
skyrmion pattern predicted in a two-component BEC, in which one component,
which carries the vorticity, is effectiviely trapped in a doughnut region
created by the other component, the entire configuration being stable \cite%
{skyrmion}.

Equations (\ref{psi}) and (\ref{phi}) conserve the norms of the two wave
functions, i.e., numbers of atoms in the species, $\int \left\vert \psi
_{1,2}(\mathbf{r},t)\right\vert ^{2}\,d\mathbf{r}=N_{1,2}$. The conserved
energy (Hamiltonian) of the model is
\begin{eqnarray}
E &=&\int \left\{ \frac{1}{2}|\nabla \psi _{1}|^{2}+\frac{1}{2m}|\nabla \psi
_{2}|^{2}\right.  \nonumber \\
&&+\frac{1}{2}(x^{2}+y^{2})\left[ |\psi _{1}|^{2}+\Omega ^{2}\left\vert \psi
_{2}\right\vert ^{2}\right]  \nonumber \\
&&\left. +\frac{1}{2}\left[ g_{1}|\psi _{1}|^{4}+g_{2}\left\vert \psi
_{2}\right\vert ^{4}\right] +g_{12}|\psi _{1}|^{2}|\psi _{2}|^{2}\right\} d%
\mathbf{r}\;.  \label{energy}
\end{eqnarray}%
It is well known that, in the case of repulsion between atoms, the two
species in the free space are miscible under the condition $%
g_{12}^{2}/\left( g_{1}g_{2}\right) <1 $, and miscible in the opposite case
\cite{miscibility}. The pressure induced by the transverse confinement
changes the situation, pushing the system towards stronger miscibility (see,
e.g., Ref. \cite{Ilya}).

We aim to study a different effect, \textit{viz}., a shift of the
miscibility threshold induced by vorticity imparted to one of the species.
First, we derive a system of effective one-dimensional (1D) nonpolynomial
Schr\"{o}dinger equations (NPSEs) for the rotating and nonrotating species
trapped in the pipe. The derivation extends the earlier developed analysis
of the single-component \cite{sala-npse} and two-component \cite{sala-boris}
settings, as well as the single-component case with the vorticity \cite%
{vortex-npse}. Then, the coupled NPSEs are used to predict, in an analytical
form, the threshold for the transition to the immiscibility in the mixture
including the rotating and nonrotating components. Finally, the prediction
is compared to direct numerical solutions of the 3D equations, (\ref{psi})
and (\ref{phi}), which demonstrates a high accuracy of the analytical
miscibility condition. Actual mixed and phase-separated configurations, and
the onset of suction, are presented using numerical solutions of the 3D
equations.

Coupled GPEs (\ref{psi}) and (\ref{phi}) can be derived from the action
functional,
\begin{equation}
S=\int dt\int d\mathbf{r}\left\{ -E+\frac{i}{2}\int \left( \psi _{1}^{\ast }{%
\frac{\partial }{\partial t}}\psi _{1}+\psi _{2}^{\ast }{\frac{\partial }{%
\partial t}}\psi _{2}\right) \right\} \,,  \label{action}
\end{equation}%
with $E$ taken as per Eq. (\ref{energy}). To derive effective 1D equations
in the axial direction ($z$), we generalize the 3D ansatz proposed for
nonrotating configurations in Ref. \cite{sala-boris}, assuming that the
vorticity, which is quantified by integer positive ``spin" $S $, is imparted
to the first species, while the second species has no vorticity:
\begin{equation}
\psi _{1}(\mathbf{r},t)={\frac{(x^{2}+y^{2})^{S/2}}{\sqrt{\pi S!}\sigma
_{1}(z,t)^{S+1}}}\exp \left( {iS\theta -}\frac{x^{2}+y^{2}}{2\sigma
_{1}^{2}(z,t)}\right) \,f_{1}(z,t),  \label{vario1-npse}
\end{equation}%
\begin{equation}
\psi _{2}(\mathbf{r},t)=\frac{1}{\sqrt{\pi }\sigma _{2}(z,t)}\exp \left( -%
\frac{x^{2}+y^{2}}{2\sigma _{2}^{2}(z,t)}\right) \,f_{2}(z,t),
\label{vario2-npse}
\end{equation}%
where $\sigma _{1,2}(z,t)$ are axially nonuniform transverse widths of the
two components, and $f_{1,2}(z,t)$ are the respective axial wave functions.
We substitute this ansatz in Eq. (\ref{action}) and perform the integration
over $x$ and $y$, neglecting the derivatives of $\sigma _{1,2}(z,t)$. In
this way, we arrive at an effective action functional. Extremizing this
functional with respect to $f_{1}^{\ast }(z,t)$ and $f_{2}^{\ast }(z,t)$
(the asterisk stands for the complex conjugation), we derive the coupled
time-dependent NPSEs,%
\[
i{\frac{\partial f_{1}}{\partial t}}=\left[ -{\frac{1}{2}}{\frac{\partial
^{2}}{\partial z^{2}}}+{\frac{S+1}{2}}({\frac{1}{\sigma _{1}^{2}}}+\sigma
_{1}^{2})\right.
\]%
\begin{equation}
\left. +{\frac{(2S)!}{2^{2S}(S!)^{2}}}{\frac{g_{1}|f_{1}|^{2}}{2\pi \sigma
_{1}^{2}}}+{\frac{g_{12}\,\sigma _{2}^{2S}|f_{2}|^{2}}{\pi (\sigma
_{1}^{2}+\sigma _{2}^{2})^{S+1}}}\right] f_{1}\;,  \label{npse1}
\end{equation}%
\[
i{\frac{\partial f_{2}}{\partial t}}=\left[ -{\frac{1}{2m}}{\frac{\partial
^{2}}{\partial z^{2}}}+{\frac{1}{2}}({\frac{1}{m\ \sigma _{2}^{2}}}+\Omega
^{2}\ \sigma _{2}^{2})\right.
\]%
\begin{equation}
\left. +{\frac{g_{2}|f_{2}|^{2}}{2\pi \sigma _{2}^{2}}}+{\frac{%
g_{12}\,\sigma _{2}^{2S}|f_{1}|^{2}}{\pi (\sigma _{1}^{2}+\sigma
_{2}^{2})^{S+1}}}\right] f_{2}\;.  \label{npse2}
\end{equation}%
In addition, two relations are generated by varying the effective action
functional with respect to $\sigma _{1,2}(z,t)$,
\begin{equation}
\sigma _{1}^{4}=1+{\frac{(2S)!}{(S+1)2^{2S}(S!)^{2}}}{\frac{g_{1}}{2\pi }}%
|f_{1}|^{2}+2{\frac{g_{12}\,\sigma _{1}^{4}\,\sigma _{2}^{2S}}{\pi (\sigma
_{1}^{2}+\sigma _{2}^{2})^{S+2}}}|f_{2}|^{2}\;,  \label{sigma1}
\end{equation}%
\begin{eqnarray}
\Omega ^{2}\ \sigma _{2}^{4} &=&{\frac{1}{m}}+{\frac{g_{2}}{2\pi }}%
|f_{2}|^{2}+2{\frac{g_{12}\,\sigma _{2}^{2S+4}}{\pi (\sigma _{1}^{2}+\sigma
_{2}^{2})^{S+2}}}|f_{1}|^{2}  \nonumber \\
&&-2S{\frac{g_{12}\,\sigma _{2}^{2S+2}}{\pi (\sigma _{1}^{2}+\sigma
_{2}^{2})^{S+1}}}|f_{1}|^{2}\;.  \label{sigma2}
\end{eqnarray}%
In the absence of vorticity in the first species, $S=0$, Eqs. (\ref{npse1}),
(\ref{npse2}), (\ref{sigma1}), (\ref{sigma2}) reduce to those recently
derived in Ref. \cite{sala-boris} for the nonrotating two-component mixture.

In the case of $g_{12}=0$ the two components decouple and we obtain the NPSE
for the single-component BEC under transverse harmonic confinement, carrying
axial vorticity $S$ \cite{vortex-npse}. With $S=0$, the equation reduces to
the NPSE originally derived in Ref. \cite{sala-npse}. The single-component
model with $S\neq 0$ was recently studied in Ref. \cite{vortex-npse}.

To analyze how the well-known phase-separation condition in the free space
is modified in the pipe-shaped toroidal trap (i.e., as said above, we assume
periodic boundary conditions in the direction of $z$), we consider the
following initial expression for a weak perturbation of the axially uniform
configuration (i.e., a \emph{fully mixed} one), $f_{1,2}(z)=\sqrt{n_{1,2}%
\left[ 1\pm \alpha \cos \left( {z/R}\right) \right] }$, where $n_{1}$ and $%
n_{2}$ are densities of the two components in the unperturbed state, $\alpha
$ and $\beta $ the perturbation amplitudes, and $R$ the radius of the torus.
Inserting these functions into expression (\ref{energy}), we find a
correction to the effective energy at the second order in $\alpha $ and $%
\beta $,%
\[
{\frac{E_{2}}{2\pi R}}=\left[ {\frac{n_{1}}{\left( 4R\right) ^{2}}}+{\frac{%
(2S)!}{2^{2S}(S!)^{2}}}{\frac{n_{1}^{2}g_{1}}{8\pi \sigma _{1}^{2}}}\right]
\alpha ^{2}
\]%
\[
+\left[ {\frac{n_{2}}{\left( 4R\right) ^{2}m}}+{\frac{n_{2}^{2}g_{2}}{8\pi
\sigma _{2}^{2}}}\right] \beta ^{2}-{\frac{n_{1}n_{2}g_{12}\sigma _{2}^{2S}}{%
2\pi (\sigma _{1}^{2}+\sigma _{2}^{2})^{S+1}}}\alpha \beta \;,
\]%
where widths $\sigma _{1}$ and $\sigma _{2}$ are taken as per Eqs. (\ref%
{sigma1}) and (\ref{sigma2}) with constant densities $n_{1}$ and $n_{2}$.

The uniform configuration, with $\alpha =\beta =0$, is stable against the
phase-separating axial density modulations, i.e., the binary BEC is \textit{%
miscible}, if the respective energy curvature, ${\frac{\partial ^{2}E_{2}}{%
\partial \alpha ^{2}}}{\frac{\partial ^{2}E_{2}}{\partial \beta ^{2}}}%
-\left( {\frac{\partial ^{2}E_{2}}{\partial \alpha \partial \beta }}\right)
^{2}, \label{curvature} $ is positive at $\alpha =\beta =0$. This condition
was recently applied to the investigation of the stability of various
mixtures, \textit{viz}., 3D Fermi-Fermi \cite{sala-ps-fermi} and
Bose-Fermi-Fermi ones in terms of the BCS-BEC crossover \cite{sala-flavio1},
and 1D Bose-Fermi-Fermi mixture in the Tonks-Girardeau regime \cite%
{sala-sadhan-last}.

The critical curve of the phase separation is determined by equating the
energy curvature to zero, which yields%
\[
{\frac{\sigma _{2}^{4S}}{(\sigma _{1}^{2}+\sigma _{2}^{2})^{2S+2}}}%
g_{12}^{2}={\frac{(2S)!}{2^{2S+2}(S!)^{2}\sigma _{1}^{2}\sigma _{2}^{2}}}%
g_{1}g_{2}
\]
\begin{equation}
+{\frac{\pi }{8n_{1}n_{2}R^{2}}}\left[ {\frac{(2S)!n_{2}g_{1}}{%
2^{2S}(S!)^{2}\sigma _{1}^{2}m}}+{\frac{n_{1}g_{2}}{\sigma _{2}^{2}}}+{\frac{%
\pi }{2mR^{2}}}\right] \;.  \label{strong-condition}
\end{equation}%
From this condition it follows that, even for $g_{1}=g_{2}=0$ (no
intra-species repulsion, while the inter-species repulsion is in the
action), the mixed state exists if the value of the coefficient $g_{12}$
accounting for the repulsion between the species, is below a critical value,
i.e. if:
\begin{equation}
g_{12}<g_{12}^{(c)}={\frac{\pi }{4\sqrt{mn_{1}n_{2}}R^{2}}}{\frac{(\sigma
_{1}^{2}+\sigma _{2}^{2})^{S+1}}{\sigma _{2}^{2S}}}\;.  \label{finite}
\end{equation}%
This is clearly a finite-size effect, since expression (\ref{finite})
vanishes at $R\rightarrow \infty $.

If the density is small enough, relations (\ref{sigma1}) and (\ref{sigma2})
may be approximated by $\sigma _{1}\approx 1$ and $\sigma _{2}\approx
1/(m^{1/4}\Omega ^{1/2})$, and Eq. ({\ref{strong-condition}) yields a simple
explicit result, in the limit of }$R\rightarrow \infty $, for the onset of
the "suction" effect (uniform density of the two species along $z$):{{\
\begin{equation}
{\frac{g_{12}^{2}}{g_{1}\,g_{2}}}={\frac{(2S)!}{2^{2S+2}(S!)^{2}}}{\frac{%
(m^{1/2}\Omega +1)^{2S+2}}{m^{1/2}\Omega }}.  \label{weak-condition}
\end{equation}%
In the absence of axial vorticity ($S=0$), and for the two species with the
same mass ($m=1$) and common strength of the transverse confinement ($\Omega
=1$), Eq. (\ref{weak-condition}) coincides with its counterpart for the free
(3D) space, }}$g_{12}^{2}/\left( g_{1}\,g_{2}\right) =1$.

\begin{figure}[tbp]
{\includegraphics[height=2.2in,clip]{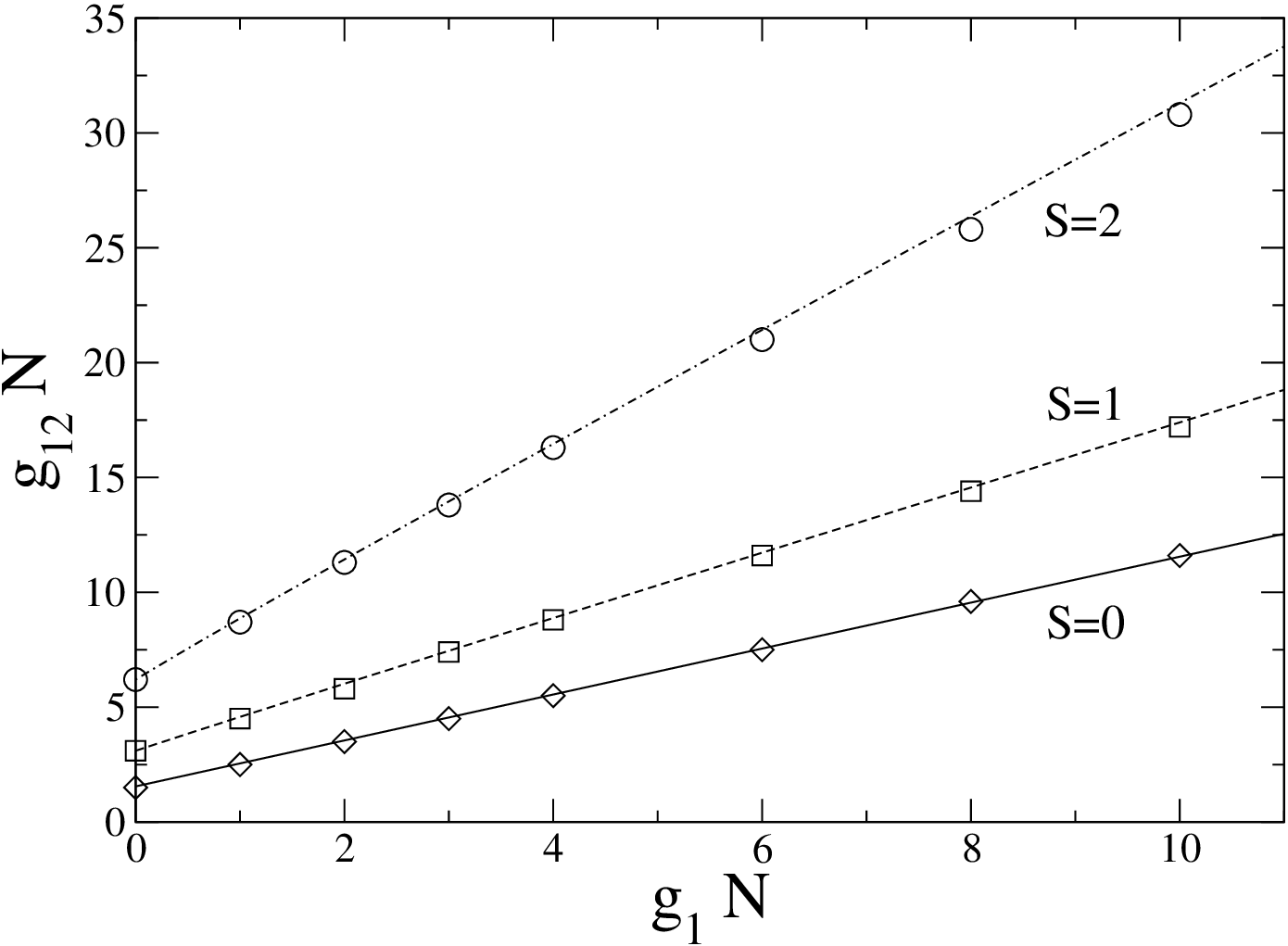}}
\caption{Critical phase-separation curves for the bosonic mixture, with $%
m=\Omega =1$, $N_{1}=N_{2}\equiv N$, and $g_{1}=g_{2}$. Lines: the
prediction of Eq. (\protect\ref{weak-condition}); symbols: results of
numerical solutions of Eqs. (\protect\ref{psi}) and (\protect\ref{phi}). }
\label{fig1}
\end{figure}

We have tested condition (\ref{weak-condition}) against direct numerical
solutions of the full 3D equations, (\ref{psi}) and (\ref{phi}). Results of
the comparison are displayed in Fig. \ref{fig1}, which shows that the
analytical criterion (\ref{weak-condition}) for phase separation is very
accurate, at least up to $g_{1}N=10$.

\begin{figure}[tbp]
\vskip 0.3cm {\includegraphics[height=2.4in,clip]{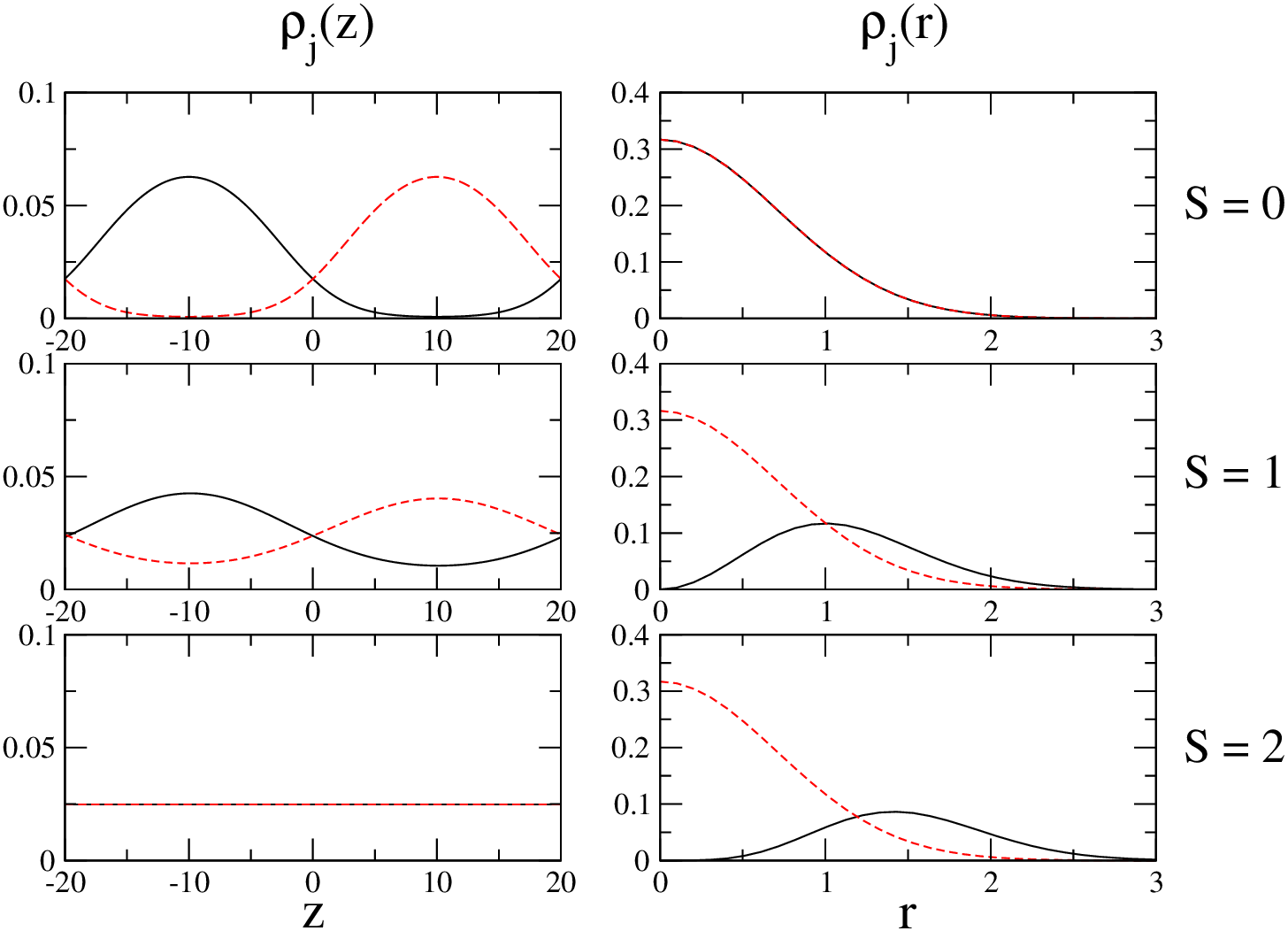}}
\caption{(Color online). The axial, $\protect\rho _{j}(z)$, and radial, $%
\protect\rho _{j}(r)$, densities [see Eqs. (\protect\ref{rho-z}) and (%
\protect\ref{rho-r})] of the two components of the binary BEC in the
quasi-1D toroidal trap. The continuous and dashed curves pertain, severally,
to the first (carrying vorticity $S$) and second (zero-vorticity)
components. The axial coordinate takes values $-20<z<+20$, with periodic
boundary conditions. Here, $m=\Omega =1$, $N_{1}=N_{2}=N$, $g_{1}N=g_{2}N=1$%
, $g_{12}N=5$.}
\label{fig2}
\end{figure}

We report also typical demixed (phase-separated) and mixed states, obtained
as direct numerical solutions of Eqs. (\ref{npse1}), (\ref{npse2}), (\ref%
{sigma1}), (\ref{sigma2}) with periodic boundary conditions along $z$, to
model the torus of a large radius. If the radius is much larger than the
transverse width $a_{\bot }$, effects of the curvature in the axial
direction may be neglected \cite{sala-ring}. We solved the equations
numerically by means of a finite-difference Crank-Nicholson
predictor-corrector method with the cylindrical symmetry (details of the
method were given in Ref. \cite{sala-numerics}). To generate the
ground-state wave functions $\psi _{1,2}(r,z)$, with $j=1,2$, the
imaginary-time integration was used.

In Fig. \ref{fig2} we plot typical examples of the axial and radial
densities,
\begin{equation}
\rho _{j}(z)=N_{j}^{-1}\int_{0}^{\infty }2\pi rdr|\psi _{1,2}(r,z)|^{2}\;,
\label{rho-z}
\end{equation}%
\begin{equation}
\rho _{j}(r)=N_{j}^{-1}\int_{-\infty }^{+\infty }dz|\psi
_{1,2}(r,z)|^{2},~j=1,2.  \label{rho-r}
\end{equation}%
Figure \ref{fig2} displays the demixing (phase separation) for $S=0$ and $%
S=1 $, while no separation is observed in the axial direction for $S=2$. In
fact, this means that the empty core induced by the vorticity with $S=2$ is
wide enough to produce the ``suction" effect, while $S=1$ is not sufficient
for that.

\begin{figure}[tbp]
{\includegraphics[height=2.2in,clip]{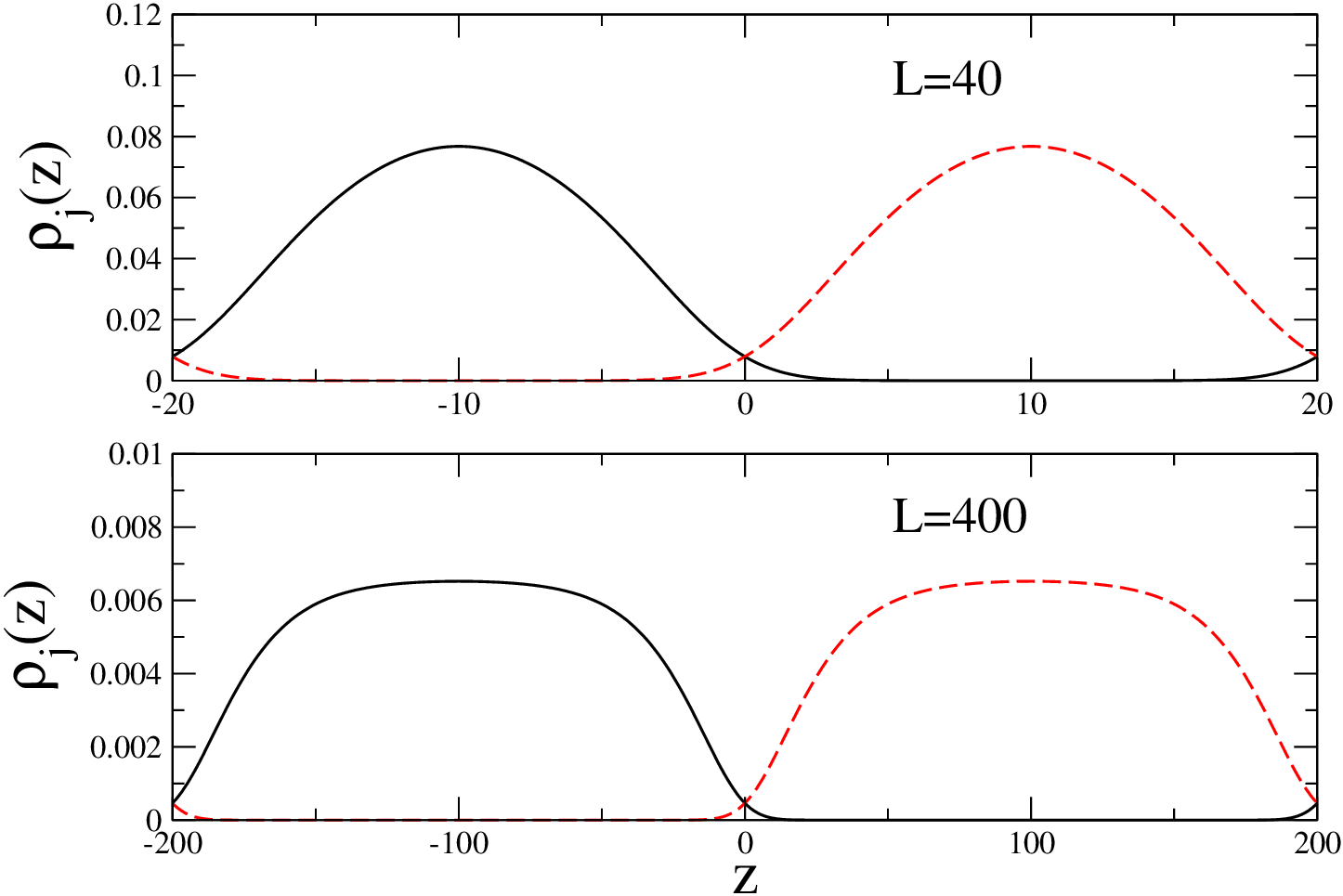}}
\caption{(Color online). The axial densities $\protect\rho _{j}(z)$ of the
two components of the binary BEC in the quasi-1D toroidal trap for two
values of the axial length, $L=40$ and $L=400$. The continuous and dashed
curves represent the first and second components. The parameters are $%
m=\Omega =1$, $N_{1}=N_{2}=N$, $g_{1}N=g_{2}N=1$, $g_{12}N=30$, $S=0$.}
\label{fig-new}
\end{figure}

To further illustrate the situation, in Fig. \ref{fig-new} we plot the axial
density profiles of the two components of the binary BEC in the regime of
the phase separation for two different values of the axial length, $L\equiv
2\pi R=40$, and $L=400$, i.e., \textquotedblleft short" and
\textquotedblleft long" tori, respectively. The figure shows that the
density profiles get flatter and interfaces steeper with the increase of $L$%
. In fact, for very large $L$ the contribution (generated by the gradient
terms) to the total energy (\ref{energy}) from to the interfaces become
negligible in comparison with the bulk terms (the Thomas-Fermi
approximation), the corresponding axial profiles looking as step functions.
In this limit, the onset of phase separation is accurately described by Eq. (%
\ref{weak-condition})

In conclusion, we have considered dynamical states in the binary BEC formed
by two species with repulsion between atoms, in the case when one species is
prepared in a vortical state, with vorticity $S=1$ or $2$, while the other
has zero vorticity. It is assumed that the condensate was loaded in a
quasi-1D toroidal trap, i.e., the corresponding equations were solved with
periodic boundary conditions in the axial direction. In this setting, we
have derived coupled 1D nonpolynomial Schr\"{o}dinger equations (NPSEs) for
the mean-field wave functions of the two components. Using these equations,
we have found the phase-separation threshold, as a condition for the onset
of the modulational instability of axially uniform states. The predictions
produced by the NPSEs were compared to results of direct simulations of the
underlying 3D equations, demonstrating a very good agreement. Further,
stable 3D states with the species mixed and separated in the axial direction
were found, the transition to the effective mixing being accounted for by
the suction, i.e., filling the empty core in the vortical component by the
nonrotating one. The onset of the suction depends on the size of the
vorticity in the first component, and on the relative strength of the
inter-species repulsion in comparison with the intrinsic repulsion in each
component.


\begin{thebibliography}{99}
\bibitem{Cornell1} M. R. Matthews \textit{et al.}, Phys. Rev. Lett. \textbf{%
83}, 2498 (1999).

\bibitem{Cornell2(filled-empty)} B. P. Anderson \textit{et al.}, Phys. Rev.
Lett. \textbf{85}, 2860 (2000).

\bibitem{reviews} A. L. Fetter and A. A. Svidzinsky, J. Phys. Cond. Matt.
\textbf{13}, R135 (2001); P. G. Kevrekidis \textit{et al.}, Mod. Phys. Lett.
B \textbf{18}, 1481 (2004); B. A. Malomed \textit{et al.}, J. Opt. B Quant.
Semics. Opt. \textbf{7}, R53 (2005).

\bibitem{multicomp} J. Ruostekoski, Phys. Rev. A \textbf{70}, 041601 (2004);
K. Kasamatsu, M. Tsubota, and M. Ueda, Int. J. Mod. Phys. B \textbf{19},
1835 (2005).

\bibitem{Cornell3} V. Schweikhard \textit{et al.}, Phys. Rev. Lett. \textbf{%
93}, 210403 (2004).

\bibitem{finland} M. M. M\"{o}tt\"{o}nen, V. Pietil\"{a}, and S. M. M.
Virtanen, Phys. Rev. Lett. \textbf{99}, 250406 (2007).

\bibitem{cigar-vortex} L. Salasnich, Laser Phys. \textbf{14}, 291 (2004); J.
A. M. Huhtam\"{a}ki, M. M\"{o}tt\"{o}nen,and S. M. M. Virtanen, Phys. Rev. A
\textbf{74}, 063619 (2006); B. A. Malomed \textit{et al.}, Phys. Lett. A
\textbf{361}, 336 (2007).

\bibitem{torus-experiment} S. Gupta \textit{et al.}, Phys. Rev. Lett.
\textbf{95}, 143201 (2005); A. S. Arnold, C. S. Garvie, and E. Riis, Phys.
Rev. A \textbf{73}, 041606(R) (2006); C. Ryu \textit{et al.},
arXiv:0709.0012.

\bibitem{torus-theory} R. Kanamoto, H. Saito, and M. Ueda, \textit{ibid.}
\textbf{73}, 033611(R) (2006); M. Modugno, C. Tozzo, and F. Dalfovo, \textit{%
ibid}. \textbf{74}, 061601(R) (2006); L. Salasnich, A. Parola, and L.
Reatto, \textit{ibid}. \textbf{74}, 036603(R) (2006); A. V. Carpentier and
H. Michinel, Europhys. Lett. \textbf{78}, 10002 (2007); I. Lesanovsky and W.
von Klitzing, Phys. Rev. Lett. \textbf{98}, 050401 (2007).

\bibitem{skyrmion} C. M. Savage and J. Ruostekoski, Phys. Rev. Lett. \textbf{%
91}, 010403 (2003).

\bibitem{miscibility} V. P. Mineev, Zh. Eksp. Teor. Fiz. \textbf{67}, 263
(1974) [Sov. Phys. JETP \textbf{40}, 132 (1974)].

\bibitem{Ilya} M. I. Merhasin, B. A. Malomed, and R. Driben, J. Phys. B
\textbf{38}, 877 (2005).

\bibitem{sala-npse} Salasnich, A. Parola, and L. Reatto, Phys. Rev. A
\textbf{65}, 043614 (2002).

\bibitem{sala-boris} L. Salasnich and B. Malomed, Phys. Rev. A \textbf{74},
053610 (2006).

\bibitem{vortex-npse} L. Salasnich, B. Malomed and F. Toigo, Phys. Rev. A
\textbf{76}, 063614 (2007).

\bibitem{sala-ps-fermi} L. Salasnich \textit{et al.}, J. Phys. B: At. Mol.
Opt. Phys. \textbf{33}, 3943 (2000).

\bibitem{sala-flavio1} L. Salasnich and F. Toigo, Phys. Rev. A \textbf{75},
013623 (2007).

\bibitem{sala-sadhan-last} S. K. Adhikari and L. Salasnich, Phys. Rev. A
\textbf{76}, 023612 (2007).

\bibitem{sala-ring} A. Parola \textit{et al.}, Phys. Rev. A \textbf{72},
063612 (2005).

\bibitem{sala-numerics} E. Cerboneschi \textit{et al.}, Phys. Lett. A
\textbf{249}, 495 (1998).
\end{thebibliography}
\end{document}